# Network-Security Informed Offer-Making of Aggregator with Utility-Owned Storage Lease Opportunity: Stochastic Stackelberg Game and Distributed Solution Methods

Congcong Liu, *Student Member, IEEE*, Zhengshuo Li, *Senior Member, IEEE*

*Abstract*—Aggregators of distributed energy resources are increasingly encouraged to participate in wholesale market bidding. However, the delivery of the power they are awarded can result in over-voltage or congestion issues within the distribution network (DN). The opportunity to lease energy storage from the utility that manages the DN provides the aggregator with a means to mitigate these issues, while also benefiting the utility in terms of additional lease revenue. Nevertheless, this leasing opportunity considerably complicates the aggregator's offer-making process, as it requires the consideration of market uncertainties, uncertain power injection at DN buses, and the strategic interactions between the aggregator and the utility. This paper presents a stochastic Stackelberg game model that effectively captures the interactions between the aggregator and the utility, ensuring DN security across all potential uncertainty scenarios. Furthermore, in light of the privacy concerns of both the aggregator and the utility, two distributed solution methods are proposed. The first method follows a traditional predict-then-optimize framework and has been validated to achieve the game equilibrium. The second method employs an end-to-end framework, which has been empirically shown to yield superior economic results. Case studies conducted on 69 and 533-bus DNs illustrate the efficacy of the proposed methods.

*Index Terms*—Aggregator, distribution network security, end-to-end, shared energy storage, stochastic Stackelberg game.

## I. Introduction

THE inadequacy of flexibility within modern power systems has led to various security challenges. To address this issue, wholesale electricity markets that seek flexible resources for the transmission network (TN), allowing aggregators of distributed energy resources (DERs) within distribution networks (DN) to participate in market bidding [1]. A notable instance of this is the real-time wholesale market in Australia [2]. However, this act has also introduced potential risks; specifically, the power that an aggregator is awarded from the market and flows through the DN may compromise the security of the DN [3].

The approaches to mitigating DN security concerns can be broadly categorized into two primary types: ex-post [4][5] and ex-ante [6]-[12] approaches. The ex-post approaches typically assume that a distribution system operator (DSO), or a utility acting in the capacity of a DSO, can directly oversee the management of the aggregator's awarded power delivery when necessary [4]. Conversely, certain ex-ante approaches, such as fixed export/import limits [6], dynamic operating limits [7], and dynamic operating envelope as outlined in [8]-[12], establish a network-security offer range (NSOR) before or concurrently with the offering process. This NSOR delineates the acceptable boundaries for power injection/withdrawal by the aggregator, thereby ensuring that any power transactions conducted within this range remain compliant with DN security constraints. However, these approaches can often restrict the aggregators' bidding power range, potentially undermining their competitiveness and profitability in the wholesale market [13].

In light of the growing prominence of sharing economies, certain utility companies have initiated the provision of a service for the leasing of utility-owned shared energy storage (SES) [14]. These utility-owned energy storage systems typically possess power ratings ranging from several megawatts to 50 megawatts and exhibit significantly rapid response times [15]. Consequently, it is evident that DER aggregators may find leasing and utilizing this SES service advantageous for enhancing their competitiveness in the wholesale market [13]. Specifically, engaging in SES leasing (hereafter referred to as the *SES mode*) can assist aggregators in avoiding potential DN security challenges, as the leased SES can be employed to absorb excess power or generate power when deficits occur [16][17]. This capability enables the aggregators to expand their bidding power range, or NSOR, thereby improving their competitive standing in the wholesale market.

However, the inclusion of the SES lease opportunity significantly complicates the offer-making process for the aggregator. This complexity arises from the necessity of accounting for the operational constraints of the aggregated DERs, the SES, and the security constraints of the DN. Notably, the SES device and the DERs may be situated at different DN buses, necessitating the incorporation of DN power flow constraints. Furthermore, the SES owner is the utility rather than the aggregator; the utility may strategically modify the leasing price—an expense for the aggregator—to optimize its revenue from the SES lease service. Consequently, the aggregator must factor this into its offer-

This work was supported in part by the National Natural Science Foundation of China under Grant 52377107. (Corresponding author: Zhengshuo Li).

The authors are with the School of Electrical Engineering, Shandong University, Jinan 250061, China (e-mail: ccliusdu@mail.sdu.edu.cn; zsli@sdu.edu.cn).

making process and devise an optimal leasing strategy to maximize expected offer revenue. This scenario introduces a strategic interaction, or *gaming*, between the aggregator and the utility, necessitating respect for the privacy concerns of both entities [1]. Additionally, for many aggregators, the wholesale market clearing prices are uncertain at the time of making offers, and similar uncertainty exists regarding power injections at specific DN buses. The strategic adjustment to the leasing price by the utility further is also uncertain from the aggregator's perspective.

In summary, the aggregator's offer-making process in the context of a utility-owned SES lease opportunity, encompasses DN security constraints, strategic interactions with the utility, and various uncertainties related to electricity pricing and power injection, rendering it a complex decision-making problem.

The conventional approach to addressing uncertainty in decision-making problems is the *predict-then-optimize* framework, wherein the aggregator forecasts unknown prices and subsequently optimizes its offer strategies. Existing literature has primarily concentrated on enhancing prediction accuracy [18]-[20]. However, this approach tends to decouple the relationship between prediction and decision-making, potentially leading to suboptimal outcomes [21]. To address this limitation, recent studies have proposed an *end-to-end* framework [22][23], which has demonstrated superior performance across a wide range of uncertain decision-making scenarios [23]. Applications of this end-to-end framework in power systems include voltage regulation [24], unit commitment [25], economic dispatch [26], and optimal scheduling [21][27]. Nevertheless, given that the SES-lease-involved offer-making problem entails strategic interactions between the utility and the aggregator, along with privacy concerns, the direct application of the existing end-to-end methodologies is insufficient.

This paper introduces a novel network-security informed offer-making method for the DER aggregator with a utility-owned SES lease opportunity. Initially, a stochastic Stackelberg game is employed to model the strategic interaction between the utility and the aggregator. Subsequently, two distributed solution methods are developed, based on the predict-then-optimize and end-to-end frameworks, respectively, to generate optimal and DN-secure offers in the face of uncertainty while safeguarding the privacy of both the aggregator and the utility. The contributions of this research are twofold:

1) The proposed stochastic Stackelberg game model effectively encapsulates the complex interactions between the aggregator and the utility, which arise from concerns related to the SES lease and DN security, in the context of uncertainties in market clearing price and power injection. This model is designed to facilitate mutually beneficial outcomes for both the aggregator and the utility, providing an optimal and comprehensive DN-secure bidding strategy for the aggregator, alongside an appropriate SES leasing price for both entities. In contrast to traditional methods that often restrict the range of bidding power, the proposed model empowers the aggregator to strategically utilize the SES lease opportunity, thereby enhancing its competitive position within the wholesale market.

2) Considering that conventional approaches to resolving a Stackelberg game model typically necessitate the disclosure of the private models of lower-level entities to the upper-level entity, we propose two alternative methods for privacy-preserving distributed solutions. These methods can be translated into *two distinct distributed and collaborative (D&C) decision-making frameworks* for the aggregator and the utility, allowing for their interactions without the necessity of revealing any private models. Specifically, the first D&C mode is assured of achieving equilibrium within the Stackelberg game model, while the second mode, which is based on an end-to-end framework, can yield superior economic results.

The structure of this paper is organized as follows: Section II provides a detailed statement of the offer-making problem. Section III introduces the stochastic Stackelberg game model involving the aggregator and the utility. Section IV outlines the two proposed distributed solution methods. Finally, Section V presents case studies, followed by conclusions in Section VI.

## II. PROBLEM DESCRIPTION

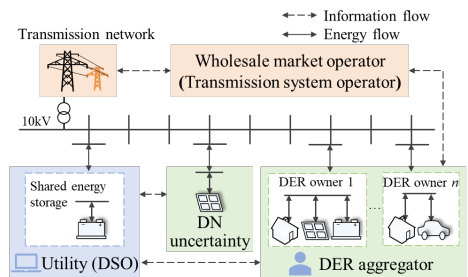

**Fig. 1.** Diagram of the interaction among the wholesale market operator, utility, and DER aggregator.

As illustrated in Fig. 1, the utility-owned SES and DERs may be connected to various DN buses. For the sake of simplification, we adopt the assumptions presented in [14] and [16], positing that the utility-owned SES is situated at the root bus. Furthermore, while some existing literature presumes that DERs managed by a single aggregator are confined to a single DN bus, we relax this assumption, permitting these DERs to be distributed across multiple buses. It is also crucial to acknowledge that uncertainty may arise in the power injection at certain buses, attributable to factors such as the stochastic nature of consumer behavior or variations in solar generation.

The involvement of a DER aggregator in the wholesale market encompasses three primary entities: the *wholesale market operator* (or transmission system operator), the *utility* (or DSO), and the *DER aggregators* themselves. The specific roles and responsibilities of these entities are outlined as follows:

---

[1]This means that the model of the DN, which is managed by the utility, should not be reported to the aggregator; meanwhile, the model of the aggregator is not reported to the utility.

- The wholesale market operator is tasked with clearing the wholesale market. Upon market closure, the operator disseminates trading outcomes, including the market clearing price and the awarded power, to all market participants.
- The utility functions as both the DSO and the SES operator. It is responsible for addressing DN security concerns based on available DN data, which encompasses power injection and withdrawal at various buses, as well as network topology and parameters. Concurrently, the utility aims to optimize its benefits derived from leasing SES services.
- The aggregator acts as a commercial and technical intermediary between the wholesale market and DER owners, operating under the assumption that it can manage the DERs through contractual agreements with their owners [9]. The aggregator is responsible for formulating offer and SES leasing strategies with the objective of maximizing profit.

Upon the market's opening, the DER aggregator submits its offer to the market operator, who subsequently clears the market and disseminates the awarded power to the aggregator. At the designated delivery time, the awarded power that flows through the DN into the TN should not compromise DN security.

It is evident that when the aggregator formulates its offer, the private models of the market operator and the utility remain unknown due to privacy considerations. Nevertheless, the aggregator, with the opportunity to lease SES, aspires to formulate an economically optimal and DN-secure offer. This issue will be addressed in subsequent sections.

### III. STOCHASTIC STACKELBERG GAME MODEL CONSIDERING SES MODE AND DN SECURITY

*A. Modeling Assumptions*

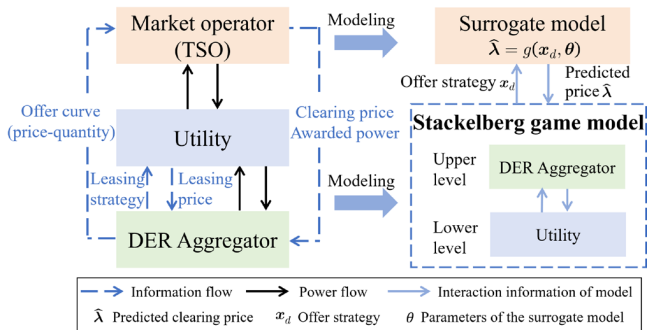

**Fig. 2.** Sketch of the proposed stochastic Stackelberg game model.

Fig. 2 illustrates the proposed stochastic Stackelberg game model. The left side of the figure depicts the interactions among the three entities discussed in Section II. The offer curve of the aggregator may influence the final clearing prices [21], yet the absence of a market operator's clearing model prevents the aggregator from accurately assessing this effect. In accordance with [22], we propose employing a *surrogate* model to approximate the relationship between the wholesale market clearing price and the aggregator's offer, stated as follows.

*Assumption 1*: The influence of the aggregator's offer on the market clearing price can be approximated through a data-driven surrogate model. The training process for this surrogate model utilizing historical data is detailed in Section IV.B.

Then, the incorporation of the surrogate model enables a concentrated examination of the interaction between the utility and the aggregator. Given the aforementioned gaming in the SES mode and uncertainties, we further introduce a *stochastic Stackelberg game* model, as delineated in Assumption 2.

*Assumption 2*: A stochastic Stackelberg game model is proposed to characterize the strategic interactions between the utility and the aggregator. Additionally, consistent with established practices in the literature, such as [20], we designate the aggregator as the leader and the utility as the follower within this framework.

*Assumption 3 (the utility plays fair)*: Although the utility is responsible for both DN security assessments and the leasing of SES, we posit that it will **not** engage in malicious manipulation of the security assessment results to coerce aggregators into leasing SES that should otherwise be avoided. Given that the utility is typically subject to governmental oversight, we believe that this fair-play assumption is likely to hold in most scenarios. The case of unfair play will be addressed in future research.

Finally, we further simplify the constraints related to the DN power flow and security as follows.

*Assumption 4*: We assume that the DN operates under approximately three-phase balanced conditions, and according to [28] that shows that in most cases "the nonlinear terms are about $10^4$ smaller compared to the linear terms", a classical linearized DistFlow model [29] is adopted in this work.

However, as will be shown in Section III.D, our Stackelberg game model requires only that the constraints be linear, thereby ensuring the applicability of the distributed solution methods in Section IV. Consequently, our work is compatible with any linear three-phase balanced or unbalanced power flow models. Given that the pursuit of high linearization accuracy has been extensively explored in the literature with favorable results [28], we will not focus on this aspect in this study.

The subsequent sections will first elaborate on the detailed stochastic Stackelberg game model, followed by a concise formulation. Throughout the remainder, a $T$-time interval horizon decision-making problem is considered where the time interval index $t \in \{1, \ldots, T\}$, and the time interval is presumed to be one time unit for the sake of simplicity.

*B. Upper Level: High-Dimensional Offer Decision Model of Aggregator*

In the upper-level optimization problem, the aggregator $b$ should optimize its offer strategy based on the market clearing price and leasing price of the SES.

*1) Objective function*

As explained previously, the clearing price is unknown to the aggregator. Considering the inevitable prediction errors, we deem its prediction value, denoted as $\tilde{\lambda}_t$, as an uncertain parameter described with the uncertainty set $\mathcal{M} = \{\tilde{\lambda} | \lambda_t^{ex} - \Delta_{mp} \leq \tilde{\lambda}_t \leq \lambda_t^{ex} + \Delta_{mp}\}$, where $\lambda_t^{ex}$ and $\Delta_{mp}$ are the expected market clearing price and the possible deviation at time interval $t$, respectively.

The SES leasing cost is divided into two parts. The first is the energy and power capacity leasing cost [30], which can be strategically adjusted by the utility [31]. Their unit prices are denoted by $\lambda_E, \lambda_P$, respectively. The second is the operation and management (O&M) cost of the SES, which is related to the charging and discharging profiles required by the aggregator $b$. The unit O&M cost $c_{OM}$ is typically considered as a constant [30].

The high-dimensional offer[2] of the aggregator $b$ is usually a series of price-quantity pairs, including the offering price, denoted as $\alpha_{b,s,t}$, and awarded power, denoted as $P_{b,s,t}$.

Then, the objective function of the aggregator $b$ that maximizes its profit, i.e., the income from bidding in the wholesale market minus the cost of leasing the SES and its own operational cost, is formulated as follows:

$$\max_{\Phi} \min_{\tilde{\lambda}_t \in \mathcal{S}} \sum_t \sum_s (\tilde{\lambda}_t P_{b,s,t} - c_{b,s,t} P_{b,s,t}) - \lambda_E E_b^{\max} \\ - \lambda_P P_{ES,b}^{\max} - \sum_t c_{OM}(P_{ES,b,t}^c + P_{ES,b,t}^d) \quad (1)$$

where the variables set $\Phi$ includes $\{P_{b,s,t}, \alpha_{b,s,t}, E_b^{\max}, P_{ES,b}^{\max}\}$; $c_{b,s,t}$ is the unit generation cost for the aggregator $b$'s $s$-th offer pair, respectively, at time interval $t$; $E_b^{\max}$ and $P_{ES,b}^{\max}$ are the energy and power capacities leased by aggregator $b$, respectively; $P_{ES,b,t}^c$ and $P_{ES,b,t}^d$ are the charging and discharging quantity of the leased SES leveraged by aggregator $b$, respectively, at time interval $t$.

The first item in (1) means the profit from bidding in the wholesale market; the second and third are the SES leasing cost; and the last term is the SES O&M cost.

Without loss of generality, objective (1) can be equivalently reformulated as (2) [33]:

$$\max_{\Phi,z} z - \sum_t \sum_s c_{b,s,t} P_{b,s,t} - \lambda_E E_b^{\max} - \lambda_P P_{ES,b}^{\max} \\ - \sum_t c_{OM}(P_{ES,b,t}^c + P_{ES,b,t}^d), \quad (2)$$

and the auxiliary variable $z \leq \sum_t \sum_s \tilde{\lambda}_t P_{b,s,t}, \forall \tilde{\lambda}_t \in \mathcal{M}$.

*2) Operational model of leased SES*

The constraints related to the leased SES are as follows:

$$0 \leq P_{ES,b,t}^c \leq P_{ES,b}^{\max}, 0 \leq P_{ES,b,t}^d \leq P_{ES,b}^{\max} \quad \forall t \quad (3)$$

$$P_{ES,b,t}^c \leq k_c E_b^{\max}, P_{ES,b,t}^d \leq k_d E_b^{\max} \quad \forall t \quad (4)$$

$$P_{ES,b,t} = P_{ES,b,t}^d - P_{ES,b,t}^c \quad \forall t \quad (5)$$

$$E_{b,t} = E_{b,t-1} + \eta_+ P_{ES,b,t}^c - P_{ES,b,t}^d / \eta_- \quad \forall t \quad (6)$$

$$E_b^{\min} \leq E_{b,t} \leq E_b^{\max} \quad \forall t, \quad E_{b,0} = E_{b,T} \quad (7)$$

$$0 \leq E_b^{\max} \leq E^{\max}, 0 \leq P_{ES,b}^{\max} \leq P_{ES}^{\max} \quad (8)$$

$$P_{ES,b}^{\max} \leq k E_b^{\max} \quad (9)$$

[2] In electricity markets, a high-dimensional offer is defined as a set of price-quantity pairs arranged in a monotonically increasing order. The term "high-dimensional" means the dimension of the offer, which is the sum of the price and quantity dimensions [32].

where $k_c$, $k_d$, and $k$ are the energy storage charging, discharging, and leasing limiting coefficients, respectively; $E_{b,t}, E_{b,0}$, and $E_{b,T}$ are the energy at time interval $t$, initial, and the final energy state of the leased SES, respectively; $\eta_+$ and $\eta_-$ are the charging and discharging efficiencies, respectively; and $E^{\max}$ and $P_{ES}^{\max}$ are the energy and power capacities of the SES, respectively.

Constraint (3) is the upper and lower limits of the charging and discharging power, and do not exceed the power capacity of the leased SES. Constraint (4) limits the charging and discharging power to not exceed the power limit of the leased SES, which is typically proportional to the energy capacity. Constraint (5) represents the power output of the leased SES. Constraint (6) describes the energy dynamics. Constraint (7) is the capacity limit of the leased SES for preventing overcharging and over-discharging, and limits the initial energy to the same value as that at the final energy. We assume that the initial energy is half of the energy capacity and is supplied by the utility [30]. Constraints (8) and (9) represent the energy capacity, power capacity, and proportionality constraints of the leased SES, respectively.

*3) Constraints for aggregate power from DERs*

Let $P_{Ag,b,t}$ denote the aggregate power from DERs of aggregator $b$, including distributed generation $P_{DG,j,b,t}$, demand response $P_{DR,j,b,t}$, and distributed energy storage $P_{DES,j,b,t}$, where the subscript $j$ denotes the connection bus.

$$P_{Ag,b,t} = \sum_j (P_{DG,j,b,t} + P_{DR,j,b,t} + P_{DES,j,b,t}) \quad \forall t \quad (10)$$

$$P_{DG,j,b,t} \in \Omega_{DG,j,b,t}, P_{DR,j,b,t} \in \Omega_{DR,j,b,t}, \\ P_{DES,j,b,t} \in \Omega_{DES,j,b,t} \quad \forall t, \quad (11)$$

where $\Omega_{DG,j,b,t}, \Omega_{DR,j,b,t}$, and $\Omega_{DES,j,b,t}$ are the sets of the operational constraints of $P_{DG,j,b,t}, P_{DES,j,b,t}$, and $P_{DR,j,b,t}$ at bus $j$, respectively. $\Omega_{DG,j,b,t}, \Omega_{DR,j,b,t}$ include the upper and lower limits of power. $\Omega_{DES,j,b,t}$ includes the limits of power and energy, and the state of charge constraints. The detailed formula can be referred to [9][34].

The generation cost of the aggregator $b$ is typically considered equivalent to the cost of the aggregated DERs:

$$\sum_s c_{b,s,t} P_{b,s,t} = \sum_j (c_{DG,b} P_{DG,j,b,t} \\ + c_{DR,b} P_{DR,j,b,t} + c_{DES,b} P_{DES,j,b,t}) \quad \forall t \quad (12)$$

where $c_{DG,b}, c_{DR,b}$, and $c_{DES,b}$ represent the cost coefficients associated with distributed generation, demand response, and distributed energy storage, respectively.

*4) Constraints for high-dimensional offer curve*

Typically, the aggregator's offer curve should be incremental and higher than its operational cost at the same pair:

$$\alpha_{b,s,t} \leq \alpha_{b,s+1,t}, \alpha^{\min} \leq \alpha_{b,s,t} \leq \alpha^{\max} \quad \forall s, \forall t \quad (13)$$

$$\alpha_{b,s,t} \geq c_{b,s,t} \quad \forall s, \forall t \quad (14)$$

where $\alpha^{\max}$ and $\alpha^{\min}$ are the upper and lower limits of the offer price, respectively, determined by market rules.

The awarded power of the aggregator's $s$-th offer pair should

be within the pair capacity limit $[P_{b,s,t}^{\min}, P_{b,s,t}^{\max}]$:

$$P_{b,s,t}^{\min} \leq P_{b,s,t} \leq P_{b,s,t}^{\max} \quad \forall s, \forall t \tag{15}$$

The awarded power $P_{g,b,t}$ of the aggregator $b$ should be located within the NSOR $[P_{g,b,t}^{\min}, P_{g,b,t}^{\max}]$ and it is equal to the sum of all pairs' awarded power. Further, $P_{g,b,t}$ is the sum of the aggregate power $P_{Ag,b,t}$ and the leased SES $P_{ES,b,t}$ (when the SES is in charging state, $P_{ES,b,t}$ is negative, otherwise positive), as shown below.

$$P_{g,b,t} = \sum_s P_{b,s,t} \quad \forall t \tag{16}$$

$$P_{g,b,t}^{\min} \leq P_{g,b,t} \leq P_{g,b,t}^{\max} \quad \forall t \tag{17}$$

$$P_{g,b,t} = P_{ES,b,t} + P_{Ag,b,t} \quad \forall t \tag{18}$$

### C. Lower Level: Decision-Making Model of Utility

In the lower-level problem, the utility has two tasks: to ensure DN security in the presence of uncertain bus power injection, and to maximize its profit by setting proper SES lease price, i.e., $\lambda_E, \lambda_P$ as explained previously.

#### 1) Objective function

The objective function is as follows,

$$\max_{\lambda_E, \lambda_P} \sum_b (\lambda_E E_b^{\max} + \lambda_P P_{ES,b}^{\max}) + \sum_t \tilde{\lambda}_t P_{ES,t}$$
$$+ \sum_t \sum_b c_{OM}(P_{ES,b,t}^c + P_{ES,b,t}^d) \tag{19}$$
$$- \sum_t c_{OM}(P_{ES,t}^c + P_{ES,t}^d)$$

where $P_{ES,t}$ is the power output of the utility utilizing the remaining capacity to participate in the electricity market; and $P_{ES,t}^c$ and $P_{ES,t}^d$ are the charging and discharging quantities of the SES at time interval $t$, respectively.

The first item in (18) represents the profit from leasing energy storage to aggregators. The second means the profit from bidding in the market by utilizing the remaining capacity. The third and fourth are the O&M fee obtained from the aggregators and the actual O&M cost of the SES.

The SES is usually leased to one or more aggregators. Specifically, when multiple aggregators have different charging/discharging strategies, there is a charge and discharge offset, and the offset portion of the electricity does not incur the O&M cost. Meanwhile, the utility needs to participate in the market to release this offsetting power.

#### 2) Cost constraints of the leasing prices

The leasing prices of energy and power capacity should be higher than the investment cost, satisfying:

$$\lambda_E \geq k_r c_E, \lambda_P \geq k_r c_P, \text{ and } k_r = \frac{r(1+r)^y}{365[(1+r)^y-1]} \tag{20}$$

where $k_r$ is the cost coefficient; $r$ is the discount rate; $y$ is the lifetime of the energy storage devices; and $c_E$ and $c_P$ are the unit investment costs of the energy and power capacities for the SES, respectively.

#### 3) Operational constraints of SES

The related constraints of SES are similar to the constraints (5)-(9) in the upper-level model.

$$P_{ES,t}^d - P_{ES,t}^c = \sum_b P_{ES,b,t} + P_{ES,t} \tag{21}$$

$$E_t = E_{t-1} + P_{ES,t}^c - P_{ES,t}^d - P_{Loss,t} \tag{22}$$

$$E^{\min} \leq E_t \leq E^{\max}, E_0 = E_T \tag{23}$$

$$0 \leq P_{ES,t}^c \leq P_{ES}^{\max}, 0 \leq P_{ES,t}^d \leq P_{ES}^{\max} \tag{24}$$

The power loss $P_{Loss,t}$ of the SES (25) can be transformed into (26) based on [35].

$$P_{Loss,t} = \max\{(1/\eta_- - 1)P_{ES,t}^d, (1-\eta_+)P_{ES,t}^c\} \tag{25}$$

$$P_{Loss,t} \geq (1/\eta_- - 1)P_{ES,t}^d, P_{Loss,t} \geq (1-\eta_+)P_{ES,t}^c \tag{26}$$

where $E_t$ is the loss and energy state of the SES at time interval $t$.

#### 4) DistFlow model under uncertainty

The root bus voltage and power injection $\tilde{P}_{inj,j,t}^{\text{Ori}}$ in the DN are uncertain, and the awarded power of the aggregator is not determined. Therefore, we need to model the DN power flow under uncertainty, that is, the DistFlow model under uncertainty. First, we model the DN uncertainty into a set. The uncertainty set $\tilde{u} = [[\tilde{P}_{inj,j,t}^{\text{Ori}}]; \tilde{v}_{set}; \tilde{u}_{P,b}]$ is constrained by the following polyhedron $\mathcal{U}$:

$$\mathcal{U} = \left\{ \tilde{u} \middle| \begin{array}{c} P_{inj,j,t}^{ex} - \Delta_{g,j} \leq \tilde{P}_{inj,j,t}^{\text{Ori}} \leq P_{inj,j,t}^{ex} + \Delta_{g,j} \\ v_{set}^{\min} \leq \tilde{v}_{set} \leq v_{set}^{\max}, 0 \leq \tilde{u}_{P,b} \leq 1 \end{array} \right\} \tag{27}$$

where $\tilde{P}_{g,b,t} = P_{g,b,t}^{\min} + \tilde{u}_{P,b}(P_{g,b,t}^{\max} - P_{g,b,t}^{\min})$
$$\text{for any } \tilde{u}_{P,b} \in [0,1] \tag{28}$$

where $P_{inj,j,t}^{ex}$ and $\Delta_{g,j}$ are the expected active power injection and the possible deviation at bus $j$, respectively; $\tilde{v}_{set}$ is the square of the voltage at root bus, whose range is $[v_{set}^{\min}, v_{set}^{\max}]$, which can be known through historical data; and $\tilde{u}_{P,b}$ is an uncertain variable that varies between [0,1].

Consequently, for any point $\tilde{u}$ within the uncertainty set (27)(28), the following model (29) holds. Thus, there exists a set of feasible solutions, that is, a set of power-flow calculation results within security.

$$\begin{cases} P_{inj,j,t} = \tilde{P}_{inj,j,t}^{\text{Ori}} + \sum_b (P_{DG,j,b,t} \\ \qquad\qquad + P_{DR,j,b,t} + P_{DES,j,b,t}) \\ P_{inj,j,t} = \sum_{k \in \mathcal{T}(j)} P_{jk,t} - \sum_{i \in \mathcal{H}(j)} P_{ij,t} \\ Q_{inj,j,t} = \sum_{k \in \mathcal{T}(j)} Q_{jk,t} - \sum_{i \in \mathcal{H}(j)} Q_{ij,t} \\ v_{i,t} - v_{j,t} = 2(r_{ij}P_{ij,t} + x_{ij}Q_{ij,t}) \\ v_{1,t} = \tilde{v}_{set} \\ v_j^{\min} \leq v_{j,t} \leq v_j^{\max} \quad \forall j \in \mathcal{B}/\{1\} \end{cases} \begin{array}{l} \forall \tilde{u} \in \mathcal{U}, \\ j \in \mathcal{B}, \\ (i,j) \in \mathcal{E} \end{array} \tag{29}$$

where $P_{inj,j,t}$ and $Q_{inj,j,t}$ are the total active and reactive power injection/withdrawals at bus $j$, respectively, and it is positive when injecting, otherwise negative; $\mathcal{T}(j)$ and $\mathcal{H}(j)$ are the sets of the parent and child buses of bus $j$, respectively; $P_{ij,t}$ and $Q_{ij,t}$ are the active and reactive power flowing through branch $(i,j)$, respectively; $r_{ij}$ and $x_{ij}$ are the resistance and impedance of branch $(i,j)$, respectively; and $v_{j,t}$ is the square of the voltage at bus $j$, whose operational limits are $v_j^{\max}$ and $v_j^{\min}$.

Note that, this paper focuses only on the active power range. To simplify the optimization problem, we assume that the reactive power injection/withdrawals of all buses in the above model are held constant. This assumption does not affect the conclusions reached in this study.

*D. Compact Model and All-Scenario-Security Guarantee*

We denote the optimization variables for the aggregator $b$ optimization problem (2)-(18) and the utility optimization problem (19)-(29) by $x_b$ and $y$, respectively, where $x_b = [[P_{b,s,t}]; [\alpha_{b,s,t}]; [P_{g,b,t}^{\min}]; [P_{g,b,t}^{\max}]; [P_{g,b,t}]; [P_{ES,b,t}]; [P_{Ag,b,t}];$ $E_b^{\max}; P_{ES,b}^{\max}; [P_{ES,b,t}^d]; [P_{ES,b,t}^c]; z]$, $y = [[P_{ES,t}^d]; [P_{ES,t}^c];$ $\lambda_E; \lambda_P; [P_{ES,t}]]^T$. The compact form of the above stochastic Stackelberg game model is

$$\min_{x_b} f(x_b, y) = c_1^T x_b - c_2^T x_b + y^T x_b \quad (30a)$$

$$s.t. \quad A_1 x_b \leq b, \; B x_b = d \quad (30b)$$

$$c_2^T x_b \leq \tilde{\lambda}^T x_b \quad \forall \tilde{\lambda} \in \mathcal{M} \quad (30c)$$

$$y = \underset{y}{\arg\min} \begin{cases} (c_3^T - \tilde{\lambda}^T) y - x_b^T y - c_4^T x_b \\ s.t. \; C y \leq g_1, \; D y = h_1(x_b) \\ \forall \tilde{u} \in \mathcal{U}, \exists z \begin{cases} E \tilde{u} + F z \leq g_2 \\ G \tilde{u} + H z = h_2 \end{cases} \\ \mathcal{U} = \{\tilde{u} | R \tilde{u} \leq r\} \end{cases} \quad (30d)$$

The bold uppercase letters denote the coefficient matrices, and the bold small letters denote the vectors in the constraints.

The model is a bilevel optimization problem with multiple uncertainties. The upper-level problem contains the uncertainty parameter $\tilde{\lambda}$, which includes uncertain market clearing price that is affected by the offer-decision variable. The lower level includes DN constraints under uncertainty.

Despite that the above price and DN uncertainties imply an infinite number of scenarios should be considered, the following Proposition 1 indicates that only a limited number of scenarios being involved can yield an all-scenario feasible solution, which means all-scenario DN security.

***Proposition 1***: If there exist feasible solutions to (30a)-(30d) for all selected vertex scenarios of the uncertainty set, the feasibility of the solutions to all possible realizations of the uncertainty set is then guaranteed.

The proof of Proposition 1 is similar to that in [36] and thus omitted to save space. Hence, we can select vertex scenarios $\mathcal{M}'$ and $\mathcal{U}'$ from the vertices of the uncertainty sets $\mathcal{M}$ and $\mathcal{U}$, respectively, and replace $\mathcal{M}$ with $\mathcal{M}'$ and $\mathcal{U}$ with $\mathcal{U}'$. Subsequently, (30c) and (30d) can be formulated as (30e) and (30f) respectively.

$$c_2^T x_b \leq \tilde{\lambda}^T x_b \quad \forall \tilde{\lambda} \in \mathcal{M}' \quad (30e)$$

$$y = \underset{y}{\arg\min} \begin{cases} (c_3^T - \tilde{\lambda}^T) y - x_b^T y - c_4^T x_b \\ s.t. \; C y \leq g_1, \; D y = h_1(x_b) \\ \forall \tilde{u} \in \mathcal{U}', \exists z \begin{cases} E \tilde{u} + F z \leq g_2 \\ G \tilde{u} + H z = h_2 \end{cases} \end{cases} \quad (30f)$$

The final model (30) to solve includes (30a), (30b), (30e), and (30f).

## IV. DISTRIBUTED SOLUTIONS (D&C MODES)

As for the above Stackelberg game model, conventional solution methods typically require disclosing the lower-level utility's private model to the upper-level aggregator, violating the privacy requirement of the entities. In response to this issue, this section proposes two distribution methods that can protect the privacy of all entities, and can be translated into two D&C decision-making modes for the utility and aggregator.

*A. D&C Mode 1*

In this first solution method, or *D&C mode 1*, the model (30) is decomposed into two sub-problems: the upper-level problem (30a), (30b), and (30e) related to the aggregator and the lower-level problem (30f) related to the utility. We denote $x_b^*$, $y'$ and $x_b'$, $y^*$ as the optimal solutions of the upper-level and lower-level problems, respectively. Further, we denote the set of data passed by the aggregator to utility and the utility to aggregator as $\Psi_b$ and $\Psi_u$, respectively.

Consider the sum of the objective functions of the upper-level and lower-level optimization problems in (30) as a new objective function, and then stack the constraints together. Then obtain an auxiliary model (31):

$$\min_{x_b, y} \sum_b F_{Ag,b} + F_U$$

where $F_{Ag,b} = \sum_t \sum_s c_{b,s,t} P_{b,s,t} - z$,
$$F_U = \sum_t c_{OM}(P_{ES,t}^c + P_{ES,t}^d) - \sum_t \tilde{\lambda}_t P_{ES,t}$$
$$- \sum_t \sum_b c_{OM}(P_{ES,b,t}^c + P_{ES,b,t}^d) \quad (31)$$

$$s.t. \quad P_{g,b,t} = \sum_s P_{b,s,t} \quad \forall b, \forall t : \varphi_{b,t}$$

$$P_{ES,t}^d - P_{ES,t}^c = \sum_b P_{ES,b,t} + P_{ES,t} \quad \forall t : \pi_t$$

(30b), (30e), (30f)

where $\varphi_{b,t}$ and $\pi_t$ are the dual variables of equation constraints. Let $\bar{x}_b$ and $\bar{y}$ be the optimal solution of (31).

We have the following proposition:

***Proposition 2***: If a stochastic Stackelberg game equilibrium is reached under $\Psi_b$ and $\Psi_u$ [31], we have

$$\begin{aligned} x_b^*(\Psi_b; \Psi_u) &= x_b'(\Psi_b; \Psi_u) = \bar{x}_b \\ y'(\Psi_b; \Psi_u) &= y^*(\Psi_b; \Psi_u) = \bar{y} \end{aligned} \quad (32)$$

To save space, the detailed proof of proposition 2 is given in our online supplementary material [34].

This proposition implies that the solution of the optimization model (31) can reach stochastic Stackelberg game equilibrium. Based on this, the classical alternating direction method of multipliers (ADMM) [37] or its variant can be adopted to solve the model (31), which can be seen as a distributed, collaborative, and privacy-protected decision-making mode for the aggregator and the utility.

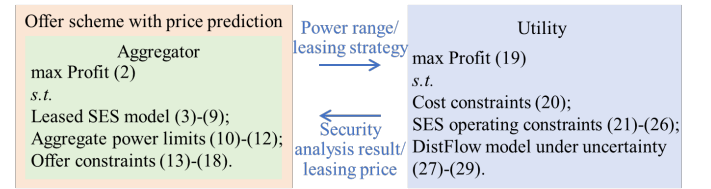

**Fig. 3.** Diagram of D&C mode 1.

The steps of this D&C mode 1 are summarized in Algorithm 1 and its diagram is shown in Fig. 3. Here, the aggregator determines an offer scheme, that is, the offer and SES leasing strategies, and submits it to the utility. The utility then analyzes the network security, updates the SES leasing price, and feeds it back to the aggregator. This interaction process is repeated until convergence is achieved.

**Algorithm 1** D&C mode 1

*Step 1:* With regard to (31), form the augmented Lagrangian function.

$$L(\boldsymbol{x}_b, \boldsymbol{y}) = \sum_b F_{Ag,b} + F_U - \sum_t \sum_b \varphi_{b,t}\left(P_{g,b,t} - \sum_s P_{b,s,t}\right)$$
$$- \sum_t \pi_t \left(P_{ES,t}^d - P_{ES,t}^c - \sum_b P_{ES,b,t} - P_{ES,t}\right)$$
$$+ \frac{\rho}{2} \sum_t \sum_b \left(P_{g,b,t} - \sum_s P_{b,s,t}\right)^2$$
$$+ \frac{\rho}{2} \sum_t \left(P_{ES,t}^d - P_{ES,t}^c - \sum_b P_{ES,b,t} - P_{ES,t}\right)^2 \quad (33)$$

*Step 2:* Set the maximum iteration number $k^{\max}$, a penalty factor $\rho$, error tolerances $\varepsilon^{pri}$, $\varepsilon^{dual}$ for the primal and dual feasibility conditions. Initialize variables $\boldsymbol{x}_b^0$, $\boldsymbol{y}^0$ and dual variables $\varphi_{b,t}^0$, $\pi_t^0$ and let $k = 1$.

*Step 3:* In the $(k+1)$-th iteration, the aggregator $b$ solves the following subproblem:

$$\boldsymbol{x}_b^{k+1} := \arg\min_{\boldsymbol{x}_b} L(\boldsymbol{x}_b^k, \boldsymbol{y}^k, \varphi_{b,t}^k, \pi_t^k) \quad (34)$$

Then, the optimal solution $\boldsymbol{x}_b^{k+1}$ can be derived and is delivered to the utility.

*Step 4:* After receiving $\boldsymbol{x}_b^{k+1}$, the utility solves the following subproblem:

$$\boldsymbol{y}^{k+1} := \arg\min_{\boldsymbol{y}} L(\boldsymbol{x}_b^{k+1}, \boldsymbol{y}^k, \varphi_{b,t}^k, \pi_t^k) \quad (35)$$

Then, the optimal solution $\boldsymbol{y}^{k+1}$ can be derived.

*Step 5:* With $\boldsymbol{x}_b^{k+1}$ and $\boldsymbol{y}^{k+1}$, the aggregator calculates $\varphi_{b,t}^{k+1}$ and $\pi_t^{k+1}$ based on the following equation.

$$\varphi_{b,t}^{k+1} := \varphi_{b,t}^k - \rho\left(P_{g,b,t}^{k+1} - \sum_s P_{b,s,t}^{k+1}\right) \quad (36)$$

$$\pi_t^{k+1} := \pi_t^k - \rho\left(P_{ES,t}^{d,k+1} - P_{ES,t}^{c,k+1} - \sum_b P_{ES,b,t}^{k+1} - P_{ES,t}^{k+1}\right) \quad (37)$$

*Step 6:* The aggregator calculates the primal and dual residuals, whose formulas are

$$\begin{cases} ||r^k||_2 = \sqrt{||\boldsymbol{\varphi}^{k+1} - \boldsymbol{\varphi}^k||_2^2 + ||\boldsymbol{\pi}^{k+1} - \boldsymbol{\pi}^k||_2^2} \\ ||s^k||_2 = ||\boldsymbol{y}^{k+1} - \boldsymbol{y}^k||_2 \end{cases} \quad (38)$$

If the residuals are enough small, i.e.,
$$||r^k||_2 \leq \varepsilon^{pri}, ||s^k||_2 \leq \varepsilon^{dual} \quad (39)$$
terminate the iteration, otherwise $k = k + 1$.

*Step 7:* If $k < k^{\max}$, go to step 3. Otherwise, stop the algorithm.

**Optimality and convergence guarantee**: With Proposition 1 and because $F_{Ag,b}$ and $F_U$ are closed, proper, and convex, model (31) is convex, so based on the property of the ADMM [37], Algorithm 1 converges to the optimal solution. Then, based on Proposition 2, this optimal solution can reach stochastic Stackelberg game equilibrium.

However, it should be noted that in this D&C mode 1, the aggregator-side problem (34) adopts a conventional predict-then-optimize to make a decision. In other words, it neglects the latent relation between the aggregator's offer decision and the clearing prices, and is likely to reduce the aggregator's benefit [21]. The following D&C mode 2 will resolve this issue by leveraging the end-to-end method.

*B. D&C Mode 2*

Drawing upon [23] and [27], this D&C mode 2 transforms the above aggregator-side problem (34) into an end-to-end framework, by introducing a surrogate model to approximate the above latent relation. The procedures are outlined as follows.

The compact form of the aggregator optimization problem in (34) can be expressed as follows:

$$f(\boldsymbol{x}_b) = \min_{\boldsymbol{x}_b} \boldsymbol{c}^T \boldsymbol{x}_b - \tilde{\boldsymbol{\lambda}}^T \boldsymbol{x}_b$$
$$s.t. \quad \boldsymbol{A}\boldsymbol{x}_b \leq \boldsymbol{b}, \boldsymbol{B}\boldsymbol{x}_b = \boldsymbol{d} \quad (40)$$

where $\tilde{\boldsymbol{\lambda}}$ is the clearing price vector, and we define $f^*(\tilde{\boldsymbol{\lambda}})$ as the optimal objective value with respect to cost vector $\tilde{\boldsymbol{\lambda}}$.

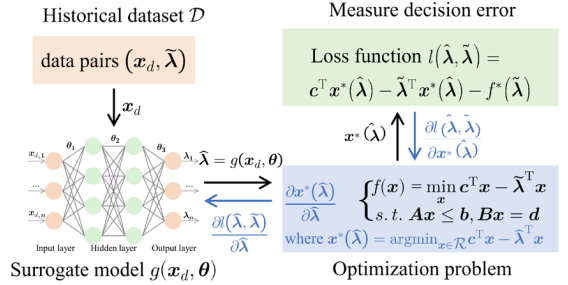

Fig. 4. End-to-end offer optimization process.

As shown in Fig. 4, the aggregator that has access to a period of historical offer strategies and clearing price data has the historical dataset $\mathcal{D} = \{(\boldsymbol{x}_{d,1}, \boldsymbol{\lambda}_1), \ldots, (\boldsymbol{x}_{d,n}, \boldsymbol{\lambda}_n)\}$, i.e., the training set. Then, the aggregator can train a surrogate model $\hat{\boldsymbol{\lambda}} = g(\boldsymbol{x}_d, \boldsymbol{\theta})$ related to the uncertain market clearing price, with the model parameters denoted by $\boldsymbol{\theta}$. Utilizing the predicted clearing price vector $\hat{\boldsymbol{\lambda}}$, the aggregator solves its optimization problem to develop an optimal offer strategy, denoted as $\boldsymbol{x}^*(\hat{\boldsymbol{\lambda}}) = \arg\min_{\boldsymbol{x} \in \mathcal{R}} \boldsymbol{c}^T \boldsymbol{x} - \hat{\boldsymbol{\lambda}}^T \boldsymbol{x}$, where $\mathcal{R}$ denotes the variable feasible region.

Moreover, the traditional predict-then-optimize framework does not consider the decision error when training the surrogate model of the clearing price. Conversely, as shown in Fig. 4, D&C mode 2, which operates within an end-to-end framework, is developed to minimize the decision error in the training process. The decision error is represented by a loss function, denoted as $l(\hat{\boldsymbol{\lambda}}, \tilde{\boldsymbol{\lambda}}) = \boldsymbol{c}^T \boldsymbol{x}^*(\hat{\boldsymbol{\lambda}}) - \tilde{\boldsymbol{\lambda}}^T \boldsymbol{x}^*(\hat{\boldsymbol{\lambda}}) - f^*(\tilde{\boldsymbol{\lambda}})$.

To integrate the optimization model with the surrogate model, the aggregator can utilize a backpropagation algorithm, which passes the gradient $\frac{\partial l(\hat{\boldsymbol{\lambda}}, \tilde{\boldsymbol{\lambda}})}{\partial \boldsymbol{\theta}}$ of the loss function back to the surrogate model to update the parameters $\boldsymbol{\theta}$.

Finally, the process is repeated until the loss function error is within the allowed range. The detailed steps are explained in Sub-algorithm 1 below.

**Sub-algorithm 1** Solution to (34) with embedded end-to-end framework

**Input**: feasible region $\mathcal{R}$, coefficient vector $\boldsymbol{c}$, historical data set $\mathcal{D}$, batch, and epochs

1: Initialize parameters $\boldsymbol{\theta}$ of prediction model
2: **for** each epoch **do**
3:    **for** each batch of training data $\mathcal{D}$ **do**
4:       Sample batch $(\boldsymbol{x}_d, \tilde{\boldsymbol{\lambda}})$
5:       $\hat{\boldsymbol{\lambda}} = g(\boldsymbol{x}_d, \boldsymbol{\theta})$
6:       $\boldsymbol{x}^*(\hat{\boldsymbol{\lambda}}) = \arg\min_{\boldsymbol{x} \in \mathcal{R}} \boldsymbol{c}^T \boldsymbol{x} - \hat{\boldsymbol{\lambda}}^T \boldsymbol{x}$
7:       $l(\hat{\boldsymbol{\lambda}}, \tilde{\boldsymbol{\lambda}}) = \boldsymbol{c}^T \boldsymbol{x}^*(\hat{\boldsymbol{\lambda}}) - \tilde{\boldsymbol{\lambda}}^T \boldsymbol{x}^*(\hat{\boldsymbol{\lambda}}) - f^*(\tilde{\boldsymbol{\lambda}})$
8:       $\frac{\partial l(\hat{\boldsymbol{\lambda}}, \tilde{\boldsymbol{\lambda}})}{\partial \boldsymbol{\theta}} = \frac{\partial l(\hat{\boldsymbol{\lambda}}, \tilde{\boldsymbol{\lambda}})}{\partial \hat{\boldsymbol{\lambda}}} \frac{\partial \hat{\boldsymbol{\lambda}}}{\partial \boldsymbol{\theta}} = \frac{\partial l(\hat{\boldsymbol{\lambda}}, \tilde{\boldsymbol{\lambda}})}{\partial \boldsymbol{x}^*(\hat{\boldsymbol{\lambda}})} \frac{\partial \boldsymbol{x}^*(\hat{\boldsymbol{\lambda}})}{\partial \hat{\boldsymbol{\lambda}}} \frac{\partial \hat{\boldsymbol{\lambda}}}{\partial \boldsymbol{\theta}} = \frac{\partial l(\hat{\boldsymbol{\lambda}}, \tilde{\boldsymbol{\lambda}})}{\partial \boldsymbol{x}^*(\hat{\boldsymbol{\lambda}})} \frac{\partial \boldsymbol{x}^*(\hat{\boldsymbol{\lambda}})}{\partial \hat{\boldsymbol{\lambda}}} \frac{\partial g(\boldsymbol{x}_d, \boldsymbol{\theta})}{\partial \boldsymbol{\theta}}$
9:       Update parameters $\boldsymbol{\theta}$ with the gradient
10:    **end for**
11: **end for**

**End-to-end embedded D&C offering-making mode:** Its algorithm is similar to Algorithm 1 except that step 3 is replaced with Sub-algorithm 1 to consider the impact of offer decisions on the forecasted clearing price. The remaining steps are performed to obtain the optimal solution of (31). However, it should be noted that since the forecasted clearing price with a decision error cannot be formulated explicitly, it is difficult to provide rigorous proof of the convergence of this D&C mode 2, and we verify its validity through case studies.

Lastly, since the focus of this study is on the decision-making of the aggregator, we assume that the utility's decision-making still adopts a traditional predict-then-optimize framework to simplify the subsequent implementation and tests. In principle, the decision-making of the utility can also be replaced with an end-to-end framework, which will be left for future study.

## V. Case Studies

To illustrate the effectiveness of the proposed approach, case studies are performed on 69-bus [10] and modified 533-bus DNs [38]. In the 69-bus DN, there are 200 end users dispersed at buses 50, 58-65. Each user at buses 58-65 is installed with a 5 kW rooftop PV and 5 kW/10 kWh battery, whose round-trip efficiency is $\eta_+ \eta_- = 85\%$ [10]. The user at bus 50 was installed with only a 5 kW rooftop PV. The aggregator aggregates the demand for buses 58-65 with all the end users, where the minimum and maximum demands were 70% and 150% of the forecasted demand, respectively. We use anonymized solar and demand data from consumers in Tasmania, Australia [3], applying the same generation on a per-capacity basis. The variation range of the root bus voltage of the DN is [0.99,1.01] p.u., and the variation range of the uncertainty of the PV output is $\pm 20\%$. When the power output of the aggregator is *negative/positive*, the aggregator functions as a *demand/generation* source, and energy flows from the grid to the aggregator or reversely. The demands of the other buses are uncontrollable.

We assume that the operational cost coefficients of the PV generation and energy storage systems are zero and 0.01 $/kWh [30], respectively. The clearing price for the PJM market in May 2022 is selected as the training set. The predicted price and initial offer range are shown in Fig. 5. The SES owned by the utility is a 10 MW/20 MWh battery whose round-trip efficiency is $\eta_+ \eta_- = 85\%$. It is located in the root bus of DN, and its O&M cost coefficient is 0.01 $/kWh. The penalty parameter $\rho$ is set as 0.01. Finally, we use the Gurobi solvers in MATLAB 2020a and Python 3.7 to solve the optimization problem.

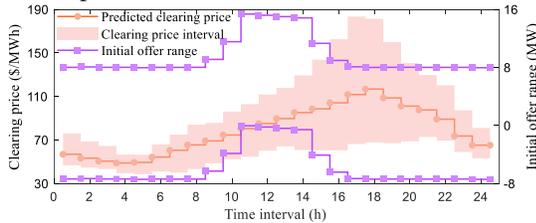

**Fig. 5.** Predicted clearing price interval and the initial offer range of aggregator.

### A. Impact of SES Mode on the Aggregator's Trading Results

To verify the effect of the SES mode on the trading results of the aggregator, we simulate the following two cases:
- *CASE A1*: The aggregator optimizes the offer strategy by solving the DN-security informed offer-making model (30). The model is solved using D&C mode 1.
- *CASE A2*: Similar to A1 except that the SES lease opportunity is absent.

Table I presents the trading results without and with SES mode. It can be seen that in CASE A1 with SES mode, the profit of the aggregator increases, where the aggregator earns $641.80 more by leasing SES. Meanwhile, the utility earns $1264.11 in revenue, which is the SES leasing fee paid by the aggregator. This indicates that the SES mode can improve the profits of both entities, namely a *mutually beneficial* outcome.

TABLE I
RESULTS OF CASES A1-A2

| Case | CASE A1 | CASE A2 |
|---|---|---|
| Profit ($) | 5512.04 | 4870.24 |
| Energy trading (MWh) | 51.80 | 54.73 |
| Energy sold (MWh) | 72.81 | 56.22 |
| Energy bought (MWh) | 21.01 | 1.49 |
| Leasing energy capacity (MWh) | 20 | - |
| Leasing power capacity (MW) | 8.68 | - |
| Leasing cost ($) | 1264.11 | - |

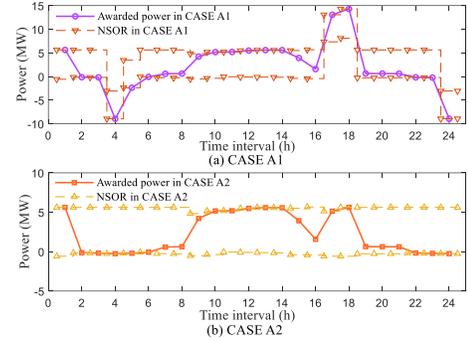

**Fig. 6.** Aggregator's awarded power and its NSOR without and with SES.

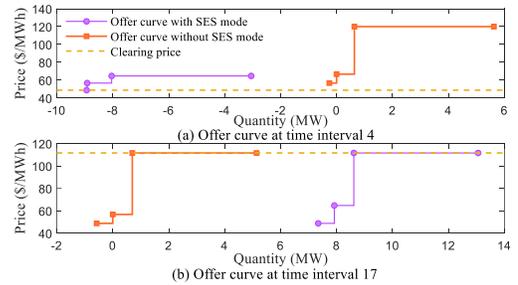

**Fig. 7.** Aggregator's offer curve at time intervals 4 and 17.

Fig. 6 and Fig. 7 further illustrate the awarded power and the offer curve related to a SES lease opportunity. In both CASE A1 and A2, the optimized NSORs are distinct from the initial one shown in Fig. 5. This divergence arises due to the incorporation of DN security constraints in both cases, which is also illustrated in Fig. 8. Here, during time interval 17, with the initial offer range, the DN faces a voltage violation issue for a certain realization of uncertain power injection. This security concern is effectively addressed in CASE A1 and A2.

Then, it is evident that in CASE A1, the SES mode enhances the aggregator's ability to bid in the wholesale market with greater flexibility: transition between a generator and a demand source according to the market clearing price profile. In particular, the NSOR and awarded power in CASE A1 are negative for time interval 4, as the SES is charged then. This facilitates the aggregator earning more profits during time interval 17 by discharging the SES at this high-price period. In a nutshell, an aggregator can leverage the SES lease opportunity to enhance its competitiveness in high-price situations.

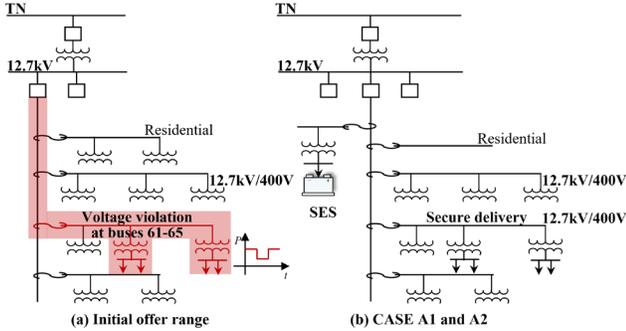

**Fig. 8.** The delivery of the awarded power during time interval 17; the similar condition also occurs for the intervals 1, 5, 11-14, 18, and 24.

### B. Comparison of Different D&C Offering-Making Modes

This part considers the following three cases on a 69-bus DN to compare two D&C offer-making modes. Here, the historical clearing price is generated based on a common wholesale market clearing program [39].

- *CASE B1*: Solve (30) using D&C mode 1. In this model, the price interval is obtained via the historical data.
- *CASE B2:* Solve (30) using D&C mode 2. The hyperparameters for training the surrogate model include a learning rate of 0.01 and a batch size of 32.
- *CASE B3*: Similar to CASE B2, but the model is solved by a centralized method in [39].

TABLE II
RESULTS OF CASES B1-B3

| Case | CASE B1 | CASE B2 | CASE B3 |
|---|---|---|---|
| Profit ($) | 3250.71 | 3258.29 | 3258.30 |
| Energy trading (MWh) | 50.56 | 49.04 | 49.04 |
| Energy sold (MWh) | 80.20 | 87.49 | 87.49 |
| Energy bought (MWh) | 29.86 | 38.45 | 38.45 |

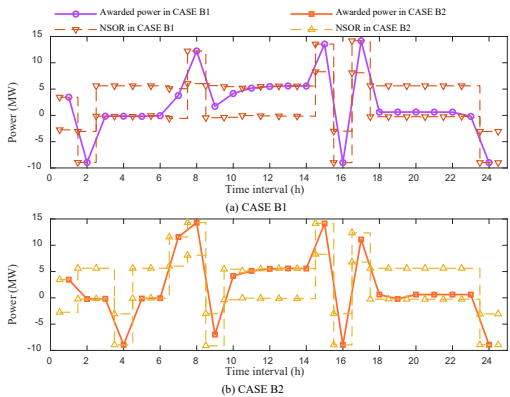

**Fig. 9.** Awarded power and NSOR in cases B1-B2.

Table II shows the trading results, indicating that CASE B2 yields higher profitability than CASE B1. The reason is that D&C mode 2 employs an end-to-end framework that incorporates the impact of the aggregator's offer strategy on the electricity price. This mode enables the optimization of the offer strategy through a more accurate prediction of the clearing price. In contrast, D&C mode 1 adopts the robust approach to address the price uncertainty, which results in a conservative offer range. This is particularly noticeable in the NSOR at time intervals 8 and 9, as depicted in Fig. 9. Therefore, optimizing the offer strategy with an end-to-end framework can enhance profitability.

Finally, a comparison between CASE B2 and B3 reveals that the trading results are nearly identical, with a variance of less than 0.1%. This finding supports the claim that the proposed D&C mode 2 achieves convergence and optimality that is comparable to the traditional centralized solution method.

### C. Solution Efficiency of two D&C Offer-Making Modes

To validate the solution efficiency of the two D&C modes, both modes are evaluated on the 69-bus and modified 533-bus DN for a 24-time interval horizon decision-making problem. In the 533-bus DN, 120 end users dispersed at buses 50, 133-135, 386-387, and 522-524. The optimality gap is defined as the percentage of the optimality value of the D&C mode with regard to that obtained by the centralized method [36].

The solution efficiency results are summarized in Table III. The optimality gap of D&C mode 2 is less than 0.01%, but that of D&C mode 1 is slightly larger.

For the 69-bus DN, the computation time of D&C mode 1 is 17.34 seconds; D&C mode 2 requires 236.71 seconds, including 224 seconds to train the parameters of the surrogate model. However, it should be noted that the computation time exhibits a slight increase with the increasing size of the DN.

In summary, D&C mode 1 has a shorter solution time, while D&C mode 2 is more economic and appropriate for scenarios where the computation time requirement is less stringent, e.g., offer in the day-ahead market. Moreover, both D&C modes preserve data privacy for both the aggregator and the utility by eliminating the need for data sharing between two entities.

TABLE III
COMPUTATIONAL PERFORMANCE OF TWO D&C MODES

| Case | 69-bus DN, D&C mode 1 | 553-bus DN, D&C mode 1 |
|---|---|---|
| Iteration number | 2 | 2 |
| Optimality gap (%) | 0.0104 | 0.0728 |
| Computation Time (second) | 17.34 | 18.10 |
| Case | 69-bus DN, D&C mode 2 | 553-bus DN, D&C mode 2 |
| Iteration number | 2 | 3 |
| Optimality gap (%) | 0.0003 | 0.0007 |
| Computation Time (second) | 236.71 | 261.02 |

## VI. CONCLUSION

This paper introduces a network-security informed offer-making method for DER aggregators participating in the wholesale market. An aggregator-utility stochastic Stackelberg game model is formulated to facilitate the dependable delivery of awarded power for aggregators amidst uncertainties related to the DN and the market clearing price. The proposed model describes the interactions between aggregators and the utility, enabling the identification of optimal strategies for both entities. Furthermore, two distributed solution modes are established to efficiently solve the model, while ensuring the data privacy of both the aggregator and the utility.

Case studies are conducted to validate that the proposed method yields a mutually beneficial outcome for the aggregator and utility. It significantly enhances the aggregator's competitiveness and increases its profit under the SES mode, while enabling the utility to earn the leasing fee. Additionally, the proposed D&C offer-making modes effectively avoid data sharing, where D&C mode 2 based on the end-to-end framework demonstrates superior economic performance with longer computation time.

Several potential directions for future research are presented below. The first is the exploration of the optimality proof of D&C mode 2 based on the end-to-end framework. The second is to extend our research to a three-phase unbalanced DN. Lastly, a more efficient training algorithm needs to be studied to reduce the training time in the D&C mode 2.